\documentclass[twocolumn,superscriptaddress,prl,amsfonts]{revtex4}
\usepackage{epsfig}

\begin{document}

\title{Finding Traps in Non-linear Spin Arrays}
\author{Marcin Wie\'sniak}\affiliation{Institut f\"ur Experimental Physik, Universit\"at Wien, Boltzmanngasse 5, A-1090 Vienna, Austria}\affiliation{Instytut Fizyki Teoretycznej i Astofizyki, Uniwersytet Gda\'nski, ul. Wita Stwosza 57, PL-80-952 Gda\'nsk, Poland}
\author{Marcin Markiewicz}\affiliation{Instytut Fizyki Teoretycznej i Astofizyki, Uniwersytet Gda\'nski, ul. Wita Stwosza 57, PL-80-952 Gda\'nsk, Poland}
\date{5.11.2009}
\begin{abstract}
Precise knowledge of the Hamiltonian of a system is a key to many of its applications. Tasks such state transfer or quantum computation have been well studied with a linear chain, but hardly with systems, which do not possess a linear structure. While this difference does not disturb the end-to-end dynamics of a single excitation, the evolution is significantly changed in other subspaces. Here we quantify the difference between a linear chain and a pseudo-chain, which have more than one spin at some site (block). We show how to estimate a number of all spins in the system and the intra-block coupling constants. We also suggest how it is possible to eliminate excitations trapped in such blocks, which may disturb the state transfer. Importantly, one uses only at-ends data and  needs to be able to put the system to either the maximally magnetized or the maximally mixed state. This can obtained by controlling a global decoherence parameter, such as temperature.
\end{abstract}
\maketitle

A very interesting application of spin chains, and more generally, lattices, is the quantum state transfer. Bose \cite{bose} noticed that having a fully magnetized state in the beginning, a spin flip propagates from the sender's site all over the system. One may then use a protocol which does not require periodic dynamics, but relies on extended infrastructure \cite{danielbose}, or a time-dependent unitary transformation \cite{bgb} to retrieve the whole information. Alternatively, one could use a system with a specific periodic evolution, which mirrors the information with respect to the middle of the chain. Such a transfer was first introduced in \cite{christ}, and independently in \cite{niko1,niko2}, and the state  mirroring condition was formalized by Shi {\em et al.} \cite{shi}.

In particular, some research has been devoted to the state transfer in the regime of limited access to the middle of the chain. This is a natural assumption, as in the microscale the quantum link between users is a black box. It was shown, for example, that with an $xx$ spin-$\frac{1}{2}$ a quantum state can be transferred without initializing the state of the interconnecting part of the chain \cite{paternostro1}, and if users can manipulate two spins, it can be even done without remote collaboration \cite{marw} (see also \cite{kay}). Interestingly, a  toy model analysis suggests that  this feature allows to better transfer information in presence of some types of decoherence \cite{marw2}. It was also shown that limited access still allows to perform a Hamiltonian tomography of whether $xx$ \cite{paternostro2} or $xxz$ chains \cite{danielkoji} and more complicated lattices \cite{danielkoji2}. It has also been shown that acting on	extreme sites of a chain one can perform quantum computation \cite{comp1,comp2}.

In this contribution we present a simple toolbox, which allows to estimate tomography of a pseudo-chain. The end-to-end evolution in the one excitation subspace (where only one spin is oriented up rather than down, OES) governed by a pseudo-chain is exactly the same as of the corresponding linear chain, but it is different in other subspaces. A pseudo-chain has a block of spins instead of a single one at some site. These spins have all equal coupling strengths to neighboring sites (blocks). Considering OES evolution, if these couplings were $J_i$, the effective coupling between blocks is $\sqrt{N_iN_{i+1}}J_i$, $N_i$ being the number of physical spins in block $i$ ($N_1=N_N=1$). An illustrative example is a map between Christandl {\em et al.}-Nikopoulos-Petrosyan-Lambropoulos chain and a hypercube of equally coupled spins described in \cite{cubes}. This might be of an interest from the physical point of view for the following reason. While some of the qubit units are complicated devices, others, such as quantum dots, are produced at random. It might be the case that in the process of fabrication not one qubit unit was placed at some point, but a number of them very close to each other. Then, if there is no possibility to directly study the structure of the chain for some reason, we provide simple tools to gain information about the system's topography.

Assume now, that we have a pseudo-chain with $xx$ interaction between nearest blocks of spins, and local magnetic fields acting on whole blocks. We start with the application of the Hamiltonian tomography by Burgarth and Maruyama \cite{danielkoji2}, who showed that working in OES we can learn all the parameters of a linear chain Hamiltonian could be learned from operating on the first site only.
 The observed dynamics would be as if the Hamiltonian was
\begin{equation}
\label{modch}
H_0=\frac{1}{2}\sum_{i=1}^N\left( J_{i}(X_{i}X_{i+1}+Y_iY_{i+1})+B'_i(1-Z_i)\right).
\end{equation}
Here with $i=1,2,...,N$ we number spin blocks (currently there is only one spin in each block), $(X_i,Y_i,Z_i)$ is the vector of Pauli matrices acting on spin in the site $i$ (except for $i=N+1$, where these operators vanish) and $J_i$'s and $B_i$'s are coupling constants, and local magnitudes of the magnetic field, respectively. The magnetic field terms have been chosen in a way that $H|00...\rangle=0$	. We will refer to the system described by (\ref{modch}) as the model chain. We assume that the actual system is described by
\begin{eqnarray}
\label{actual}
H&=&\frac{1}{2}\sum_{i=1}^N\left(\sum_{j=1}^{N_i}\sum_{k=1}^{N_{i+1}}\frac{J_i}{\sqrt{N_iN_{i+1}}}\right.\nonumber\\
&\times&(X_{i,j}X_{i+1,k}+Y_{i,j}Y_{i+1,k})\nonumber\\
&+&B_i\sum_{j=1}^{N_i}(1-Z_{i,j})\nonumber\\
&+&\left.K_{i}\sum_{j>k=1}^{N_i}(X_{i,j}X_{i,k}+Y_{i,j}Y_{i,k})\right).	
 \end{eqnarray} 
Here, $N_i$ is a number of spins-$\frac{1}{2}$ in the $i$th block, $N_1=N_N=1$ and $K_i$ represents the in-block coupling. This interaction causes a modification of the effective magnetic field observed in OES, $B'_i=B_i+(N_i-1)K_i$.

\begin{figure}
\centering
\includegraphics[width=5cm]{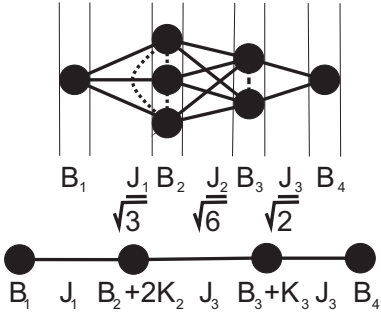}
\caption{Above: a graph representing a pseudo-chain with three spins in the second block and two in the third. Dotted and dashed lines represent intra-block couplings $K_2$ and $K_3$. Below: the respective model chain with the same end-to-end dynamics in OES.}
\end{figure}

One possible way to distinguish a pseudo-chain from a chain basing only on the at-ends data would be  to inject two excitations, one from each end. The initial state is hence $|\psi\rangle=|10...01\rangle$. Now, in a linear chain the two excitations propagate, might come close to each other, but in a pseudo-chain, they can come to the same block with more than one spin and pass each other, or then go back to the ends of the chains. The lowest number of applications of $H$ on $|\psi\rangle$ to observe this effect is $2N-2$. $N-1$ of them allow excitations to reach the block and the other half lets them go back to the end spins, where they can be observed by the users. If the passing happens at block $i$, there are $\left(\begin{array}{c}N-1\\i-1 \end{array}\right)$ ways in which the excitations reach the rendez-vous point and as many to return. If there is $N_i$ spin in the block, there are again $N_i(N_i-1)$ distinct ways for the rendez-vous to happen-first choose to which spin goes the first excitation that reaches the block, then the second.  The relative amplitude that the two excitation propagate along the chain meeting at this block is $\frac{1}{N_i^2}\prod_{j=1}^NJ_j^2$. Additionally, either of the two excitation could be the first to leave the meeting point, which produces extra 2 factor.  
Combining all these factors we reach the conclusion that
\begin{eqnarray}
\label{difference}
&\langle \psi|H^{2N-2}|\psi\rangle-\langle \psi|H^{2N-2}_0|\psi\rangle\nonumber\\
&=2\left(\prod_{i=1}^{N-1}J_i^2\right)\sum_{i=1}\left(\begin{array}{c} N-1\\i-1\end{array}\right)^2\frac{N_i-1}{N_i}.
\end{eqnarray}     

Of course, as energy is a global property, it cannot be learned directly from observing two extreme spins. However, this difference has its effects on the system evolution. Observers should deduce modified dynamics from the probability that after time $t$ the system returns to state $|\psi\rangle$,
\begin{equation}
P(|\psi\rangle\rightarrow|\psi\rangle|t)=|\langle\psi|\exp(IHt)|\psi\rangle|^2   
\end{equation}
(throughout the paper we take $\hbar=1$ and $I^2=-1$). Let us now expand the unitary evolution to the Taylor sum with respect to time. If we compare it to the evolution of $|\psi\rangle$ governed by $H_0$, we obtain that
\begin{eqnarray}
|P(|\psi\rangle\rightarrow|\psi\rangle|t)&-&P_0(|\psi\rangle\rightarrow|\psi\rangle|t)|=\frac{4}{(2N-2)!}\nonumber\\
&\times&\left|\left(\prod_{i=1}^{N-1}J_i^2\right)\sum_{i=1}\left(\frac{N_i-1}{N_i}\right)\right.\nonumber\\
&\times&\left.\left(\begin{array}{c}N-1\\i-1\end{array}\right)^2t^{2N-2}+O(t^{2N})\right|.
\end{eqnarray}
Our first procedure to reduce the class of possible geometries is as follows. First, one of the observers in turns brings the chain to the fully magnetized state, injects an excitation to find it back after some time, performing the Hamiltonian tomography \footnote{Note that the pseudo-chain is a counterexample of a system, for which driving procedures described in [D. Burgarth and V. Giovannetti, {\em Phy. Rev. Lett.} {\bf 99}, 100501 (2007)] are suitable.}. It is not necessary, but certainly beneficial if two observers do it to average the results. Basing on their data, the users can now simulate the behavior of two excitations in a linear chain, and measure their actual propagation. If they are able to identify the difference of the order of $t^{2N-2}$, they could fit the possible nets. In simple words, one can construct a table of theoretical values of the right-hand side of (\ref{difference}) in function of $N_i$'s, and choose these, which are closest to in modulo, but below the value deduced from the experiment. The underestimation of the first term in the Taylor sum follows from the fact that higher order terms in the sequence will contribute with the opposite sign, as the probability function has an oscillative character. 
Note that the procedure described above does not allow us to learn intra-block coupling constants $K_i$. Neither is it possible to order values of the pair $\{N_i,N_{N-i+1}\}$.   
These questions may be, at least partially, resolved in the other tomography routine. Burgarth and Maruyama \cite{danielkoji2} showed how to conclude coupling constants and effective local magnetic fields starting from a pure state, and this remains the first step of our procedure. Nevertheless, it was proved by Di Franco, Paternostro, and Kim \cite{paternostro2}, that for a linear $xx$ chain, this is equally possible with any initial state, in particular, the maximally mixed state. This is obviously not true for a pseudo-chain, as this feature relies on the linearity of the Hamiltonian in the fermion picture. Hence we can use this difference to determine some features of the pseudo-chain. 

In a linear chain, such as our model (\ref{modch}), the initial operator has the following form
\begin{eqnarray}
X_{1}(t)&=&e^{I H_0t}X_{1}e^{-I H_0t}\nonumber\\
&=&\alpha_{1}(t)X_{1}+\alpha_2(t)Z_{1}Y_{2}\nonumber\\
&+&\alpha_3(t)Z_{1}Z_{2}X_{3}+...\nonumber\\
&+&\beta_{1}(t)Y_{1}+\beta_2(t)Z_{1}X_{2}\nonumber\\
&+&\beta_3(t)Z_{1}Z_{2}Y_{3}+...\nonumber\\
\end{eqnarray}
with $\alpha_i=\sum_{j=0}^\infty\frac{t^j}{j!}\delta_{i,j}$, $\beta_i=\sum_{j=0}^\infty\frac{t^j}{j!}\gamma_{i,j}$, and 
\begin{eqnarray}
\delta_{i,j}&=&(-1)^{i}\left(J_{i-1}\delta_{i-1,j-1}+J_{i}\delta_{i+1,j-1}\right.\nonumber\\
&+&\left.B'_i\gamma_{i,j-1}\right),\\ \gamma_{i,j}&=&(-1)^{i+1}\left(J_{i-1}\gamma_{i-1,j-1}+J_{i}\gamma_{i+1,j-1}\right.\nonumber\\
&+&\left.B'_i\delta_{i,j-1}\right).
\end{eqnarray}
The initial conditions read $\delta_{i,0}=1$ if $i=1$ and 0 otherwise, and $\gamma_{i,0}=0$.

Assume now, that the first block containing more than one spin is labeled as $i$. Let us take $\frac{1}{2N}(1+\pm X_{1})$ as the initial state. The non-trivial part of the state reaches the block after at least $i-1$ commutations with the Hamiltonian. We now focus on the evolution of  $\left(\prod_{j<i-1}Z_{j}\right)X_{i-1}$, where we take $i$ even; when it is odd, $X_{i-1}$ is replaced with $Y_{i-1}$, but the line of the argument remains the same. Apart from propagating back to the beginning of the chain, but might also propagate through the hub to spins in the block. This part of evolution	leads to
\begin{equation}
\label{xevol1}
\left(\prod_{j<i-1}Z_{j}\right)X_{i-1}\rightarrow\frac{J_{i-1}}{\sqrt{N_{i}}}\left(\prod_{j<i-1}Z_{j}\right)\sum_{k=1}^{N_{i}}Y_{i,k}.
\end{equation}
First let us consider the effect of the magnetic field. In the model, where we have observed the effective magnetic field $B'_i$ we shall have 
\begin{eqnarray}
&\left(\prod_{j<i-1}Z_{j}\right)\sum_{k=1}^{N_{i}}Y_{i,k}&\nonumber\\
\rightarrow& -I(B_i+(N_i)K_i)\left(\prod_{j<i-1}Z_{j}\right)\sum_{k=1}^{N_{i}}X_{i,k}&,
\end{eqnarray}
which in next $i-1$ steps of evolution can retract to the beginning of the chain and contribute to $\langle X_1(0) Y_1(t)\rangle=\beta_1(t)$. When we study the actual chain, however, only the real magnetic field $B_i$ allows the operator to contribute to the auto-correlation function. Hence the we have
\begin{eqnarray}
\label{evolxtoy}
&&\frac{1}{c}|\text{Tr} Y_{1}(e^{I Ht}X_{1}e^{-I Ht}-e^{I H_0t}X_{1}e^{-I H_0t})|\nonumber\\
&=&\frac{1}{(2i-1)!}\left(\prod_{j=1}^{i-1}J_j^2\right)|(N_i-1)K_it^{2i-1}\nonumber\\
&+&O(t^{2i+1})|,
\end{eqnarray}
where the auto-correlation function is normalized to 1 at $t=0$.
  
Now, let us consider the evolution composed of $2i$ steps. There are a few trajectories such, that after the total of $2i$ commutations with the Hamiltonian the operator comes back to the form of $X_{1}$. The first is that the operator propagates back. This has a perfect analogue in the linear chain. Then it may propagate to $(i+1)$th site/block and return following a different path than the first part of propagation, which obviously impossible in the model chain. Also, the operator propagated to the block can be twice a subject to the magnetic field and the intra-block exchange. The operator might also retract to the $(i-1)$th spin and again propagate through one of $N_i$ branches of the hub. This is again in full agreement with the model evolution. The extra possibilities arise from the fact that e.g. $Z_{i-1}X_{i,1}$ does not commute with $X_{i-1}X_{i,2}+Y_{i-1}Y_{i,2}$. The propagatee gets a component, which is ``double-headed'' and proportional to 
\begin{eqnarray}
&\left(\prod_{j<i-1}Z_{j}\right)&\nonumber\\
\times&\left(Y_{i-1}\sum_{k\neq l=1}^{N_{i}}X_{i,k}X_{i,l}+X_{i-1}\sum_{k\neq l=1}^{N_{i}}Y_{i,k}X_{i,l}\right)&\nonumber.
\end{eqnarray}  	   
Then either of the two ``heads'' can be the first to retract to the origin. After analyzing all these trajectories we come to 
\begin{eqnarray}
\label{evolxtox}
&&\frac{1}{c}\left|\text{Tr}(X_{1}(e^{I Ht}X_{1}e^{-I Ht}-e^{I H_0t}X_{1}e^{-I H_0t}))\right|\nonumber\\
=&&2^N\frac{1}{(2i)!}\left(\prod_{j<i}J_j^2\right)\left(\frac{N_i-1}{N_i}\right)\nonumber\\
\times&&\left|\left(3J_{i-1}^2-J_i^2+N_iK_i\left((3N_i-2)K_i-2\sum_{j=1}^{i}B'_j\right)\right)\right.\nonumber\\ \times&&\left.t^{2i}+O(t^{2t+2})\begin{array}{c}\\ \\ \\ \end{array}\right|,\nonumber\\
\end{eqnarray}
In the bracket the first term corresponds to the scenarios with the double-headed operator, the second arises from the operator passing the block but going back the other way, and the rest originates from the difference between the real and the effective magnetic field in the block. 

The only two quantities unknown from first routine are $N_i$ and $K_i$. At most a finite set of possible solutions to Eqns. (\ref{evolxtoy}) and (\ref{evolxtox}) can be found, especially if we assume that the coupling strengths are positive. 

The second procedure to estimate the structure of the pseudo-chain is the following. Assuming that we know the effective coupling constants $J_i$ and magnetic fields $B'_i$ we simulate the dynamics governed by the model Hamiltonian (\ref{modch}). Next, we bring the system to maximally mixed state. Then we measure $X$ on the first qubit obtaining result $a=\pm 1$. After some time $t_{m,x}$ we measure $X_{1}$ getting $b_x=\pm 1$ in each run. We repeat the mixed state initialization and the measurement to reconstruct a few points of the mean function, $f_x(t_{m,x})=\langle a b_x(t_{m,x})\rangle$. We then subtract these values from those calculated for the model chain, find a polynomial in time fit to the difference and put it to Eqn. (\ref{evolxtox}). Then a similar subroutine is done, but with $Y_1$ as the second measurement and the result is used in the Eqn. (\ref{evolxtoy}). It is possible to measure $X_1+Y_1$, but then the precision of the fit is compromised, as the function must be decomposed into odd and even parts. After solving the pair of equations we might update our model by introducing the first block to it, simulate the new dynamics and by comparing with the actual data recognize the structure of next block. Finally, one can use the two-excitation procedure to verify $N_i$'s.

One should mention that because the system cannot be mapped to a free fermionic field, the state before the $X_1$ measurement cannot be arbitrary, but must be maximally mixed. One must eliminate the possibility of having an operator initially acting on spins in a block, which would then evolve to $X_1$ or $Y_1$.

A question arises if the knowledge about the defects can help us in the task of the state transfer. Consider a pseudo-chain bathed in an environment, which causes local bit flips. We assume that the coherence time is much longer than the transfer time (whether in the state-mirroring chain \cite{christ,niko1,niko2} or in the bucket scheme \cite{danielbose}). This allows the high fidelity of the transfer, but at the same time causes a difficulty, when dealing with a pseudo-chain. When the initial state was the fully magnetized one, with probability $\frac{1}{N_i}$ the excitation generated by decoherence in the $i$th block will be able leave it and reach to one the ends, where it could by taken out by one of the users (when they constantly upload the $|0\rangle$ state). But in fraction $\frac{N_i-1}{N_i}$ of all cases it is residual to the block. In this sense, blocks act as traps. It changes the magnitude of a spin by which the block could be effectively replaced and a coupling to it. The dynamics in the single excitation space becomes now an incoherent mixture of two evolutions, characteristic for two different linear chains: the one described by $H_0$ and the one effectively modified by the trapped excitation. One can now calculate such an instant of time after which the probability for the excitation injected at the beginning to be transferred to the last spin is 0, but it is sharply between 0 and 1 for the model chain. If we measure $Z_{N}$ and get $+1$ (corresponding to $|0\rangle$), we exclude neither possibility. If, however, the measurement reveals an excitation transfer at this time, it could have been only due to the evolution without the excitation trapped in block $i$. Successively, we can ``empty'' more traps in this fashion, one at the time.

In summary, we present procedures, which allow to localize additional spins in a non-linear pseudo-chain with $xx$ interaction. Importantly, our method requires only data available from measurements performed on the extreme qubits and the control over a global parameter, which we associate with temperature: at its value 0 it should bring the system to the maximally magnetized state, at infinity it introduces the maximal entropy. 
Not only do we make it possible to estimate the number of spins in each block of the system, but we are also able to tell the intra-block coupling strengths. 

We stress that our procedures reveal different features of the system than those described in \cite{danielkoji2}. Therein, the system is still studied in OES, which allows it to have ``hidden'' states, i.e. a subspace with trapped excitations. It is, however, a very complex problem to to combine our procedures with ones of \cite{danielkoji2}.

It remains an open question whether there are computation algorithms, which can be performed more efficiently with pseudo-chains than with chains. Certainly, while $xx$ chains are useless for computations, the discussed systems might find interesting applications in this direction. The reason is precisely this: non-linearity of the structure introduces interactions betweens quasi-particles. We leave these possibilities for future investigation.

Finally, we argue have argued that once we possess the knowledge about the structure of the pseudo-chains, we can remotely eliminate residual errors, which can appear due to locally acting decoherence. This allows to achieve almost perfect fidelity thanks to procedures such as the bucket scheme \cite{bgb}. It is not possible without the removal, when the evolution is effectively a statistical mixture of evolutions of two different chains. 

We gratefully acknowledge Daniel Burgarth for critical comments concerning the manuscript. MW is supported by the ESE Advanced Grant for Anton Zeilinger. This work is a part of an
\"OAD/MNiSW program and has been supported by by  the European Commission,  Project QAP  (No.   015848).  


\begin{thebibliography}{99}
\bibitem{bose} S. Bose, {\em Phys. Rev. Lett.} {\bf 91}, 207901 (2003).
\bibitem{danielbose} D. Burgarth and S. Bose, {\em Phys. Rev. A} {\bf 71}, 052315 (2005).
\bibitem{bgb} D. Burgarth, V. Giovannetti, and S. Bose, {\em Phys. Rev. A} {\bf 75}, 062327 (2007).
\bibitem{christ} M. Christandl, N. Datta, A. Ekert, and A.J. Landahl, {\em Phys. Rev. Lett.} {\bf 92}, 187902 (2004).
\bibitem{niko1} G. M. Nikolopoulos, D. Petrosyan, and P. Lambropoulos, {\em Europhys.
Lett.} {\bf 65}, 297 (2004).
\bibitem{niko2} G. M. Nikolopoulos, D. Petrosyan, and P. Lambropoulos, {\em J.
Phys.: Condens. Matter} {\bf 16}, 4991 (2004).
\bibitem{shi} T. Shi, Y. Li, Z. Song, and C. P. Sun, {\em Phys. Rev. A} {\bf 71}, 032309 (2005).
\bibitem{paternostro1} C. Di Franco, M. Paternostro, and S. M. Kim, {\em Phys. Rev. Lett.} {\bf 101} 230502 (2008).
\bibitem{marw} M. Markiewicz and M. Wie\'sniak, {\em Phys. Rev. A} {\bf 79}, 054304 (2009).
\bibitem{kay} A. Kay, arXiv:0903.4274 (quant-ph). 
\bibitem{marw2} M. Markiewicz and M. Wie\'sniak, arXiv:0905.0387 (quant-ph).
\bibitem{paternostro2} C. Di Franco, M. Paternostro, and S. M. Kim, {\em Phys. Rev. Lett.} {\bf 102} 187203 (2009).
\bibitem{danielkoji} D. Burgarth, K. Maruyama, and F. Nori, {\em Phys. Rev. A} {\bf 79} {020505(R)} (2009).
\bibitem{danielkoji2} D. Burgarth and K. Maruyama, {\em New Jour. Phys.} {\bf 11} {103019} (2009). 
\bibitem{comp1} D. Burgarth, K. Maruyama, M. Murphy, S. Montangero, T. Calarco, F. Nori, and M. B. Plenio, arXiv:0905.3373 (quant-ph).
\bibitem{comp2} A Kay and P. J. Pemberton-Ross, arXiv:0905.4070 (quant-ph).

\bibitem{cubes} M. Christandl, N. Datta, C. Dorlas, A. Ekert, A. Kay, and A. Landahl, {\em Phys. Rev. A} {\bf 71}, 032312 (2005). 

\end{thebibliography}
\end{document}